\documentstyle[12pt]{article}

\newcommand{\be}{\begin{equation}}
\newcommand{\ee}{\end{equation}}
\newcommand{\bq}{\begin{eqnarray}}
\newcommand{\eq}{\end{eqnarray}}

\newcommand{\Sc}{Schr\"odinger\,}

\begin{document}

\baselineskip=15pt

\begin{titlepage}
\begin{center}\rightline{DTP 96/39}
\vskip1in
{\LARGE Second Order Calculations of the $O(N)$ $\sigma$-Model Laplacian}
\end{center}
\vskip0.5in
\begin{center}
{\large

Jiannis Pachos

Department of Mathematical Sciences

University of Durham

South Road

Durham, DH1 3LE, England}

{\it Jiannis.Pachos@durham.ac.uk}\\

and\\

{\large
Center for Theoretical Physics \\
Laboratory for Nuclear Science \\
and Department of Physics \\
Massachusetts Institute of Technology \\
Cambridge, Massachusetts 02139} \\

{\it pachos@ctp.mit.edu}

\end{center}
\vskip0.5in
\begin{abstract}  
\noindent
For slowly varying fields on the scale of the 
lightest mass the logarithm of the vacuum functional of
a massive quantum field theory can be expanded in terms of local 
functionals satisfying a
form of the \Sc equation, the principal ingredient of which is a 
regulated functional Laplacian.
We extend a previous work to construct the next to leading order 
terms of 
the Laplacian for the Schr\"odinger equation that acts on such local 
functionals. Like the leading order the next order is completely 
determined by imposing rotational invariance in the internal 
space together with closure of the Poincar\'e algebra.

\end{abstract}

\end{titlepage}



\section{\bf The $O(N)$ $\sigma $-model Laplacian}

We are going to study the $O(N)$ $\sigma $-model as a test model to
construct the Schr\"odinger representation for non-linear systems. This 
model 
has a very interesting structure when it is quantised. Because of its
non-linearity it gives characteristics similar to the ones we have in
Yang-Mills or the Einstein cases (see \cite{mish}).

The $O(N)$ $\sigma $-model is defined as the infinite dimensional theory of
a particle on an $N$ dimensional sphere parametrised by the function $z^\mu
(\sigma ,\tau )$ as the variable $\tau $ varies. We ask the $(\sigma ,\,\tau
)$ plane to be a Minkowski space. The action of the theory has to be a
scalar with respect to the Lorentz transformations and the
reparametrisations on the sphere. So we can choose it to be 
\begin{equation}
\label{act}S=\frac 1{2\alpha }\int d\sigma d\tau g_{\mu \nu }(\dot z^\mu
\dot z^\nu -\acute z^\mu \acute z^\nu ) 
\end{equation}
where $^{\prime }$ and $\cdot $ denote differentiations with respect $\sigma$
 and $\tau $ respectively, and $\alpha $ is a coupling constant.

We can quantise this model using the Schr\"odinger Representation 
(see \cite{Sym}-\cite{Horiguchi}) by taking the field
$z(\sigma ,\tau )$ to be diagonalised at $\tau =0$ satisfying the relation
\begin{equation}
\label{momac}
z^\mu (\sigma ,0)\Psi [z]=z^\mu (\sigma )\Psi [z] 
\end{equation}
for $\Psi $ the Schr\"odinger wave functional and its conjugate momentum $\pi
(\sigma ,\tau )$ to be at $\tau =0$
\begin{equation}
\label{mom1}\pi ^\nu (\sigma,0 )\Psi [z] =i\alpha {\bf D}_\nu (\sigma 
)\Psi [z] 
\end{equation}
so that the equal time commutation relation 
\begin{equation}
\label{co1}\left[ z^\mu (\sigma,0 ),\pi _\nu (\sigma ^{\prime },0)\right]
=i\alpha \delta _\nu ^\mu \delta (\sigma ,\sigma ^{\prime }) 
\end{equation}
is satisfied. In (\ref{mom1}) the differential operator is defined with
respect to a covariant differentiation whose meaning and structure will be
given later on. Though, as $z^\mu $ is a `scalar', ${\bf D}_\nu (\sigma )$
takes the usual functional derivative form $\delta /\delta z^\nu (\sigma )$.
From (\ref{act}) we can read the Hamiltonian
\begin{equation}
\label{hami}
H=\frac \alpha 2\Delta +\frac 1{2\alpha}\int d\sigma g_{\mu \nu }\acute z^\mu 
\acute z^\nu \end{equation}
The Laplacian given in (\ref{hami}) as
${\Delta }=
\int d\sigma \,{g}
^{\mu _1\mu _2}\,{\bf D}_{\mu _1}(\sigma ){\bf D}_{\mu _2}(\sigma)=$ $
\int d\sigma _1d\sigma _2\,{\bf g}^{\mu _1\mu _2}(\sigma _1,\sigma
_2)$ $\,{\bf D}_{\mu _1}(\sigma _1)$ ${\bf D}_{\mu _2}(\sigma _2)$
for ${\bf g}^{\mu _1\mu _2}(\sigma _1,\sigma_2)={g}^{\mu _1\mu 
_2}(z(\sigma_1))\delta(\sigma_1-\sigma_2)$, is 
not well defined because the two functional derivatives act at 
the same point $\sigma $. Also the determinant of the infinite dimensional 
metric, ${\bf g}$, 
is ill-defined as the integral on its diagonal $\sigma 
_1=\sigma _2$
gives infinity. We can get around this problem by defining the 
Laplacian to have the regulated expression
\begin{equation}
\label{lapal}{\Delta }_s=\int d\sigma _1d\sigma _2\,{\bf G}^{\mu _1\mu
_2}_s(\sigma _1,\sigma _2)\,{\bf D}_{\mu _1}(\sigma _1){\bf D}_{\mu 
_2}(\sigma
_2).
\end{equation}
The Kernel ${\bf G}$, which takes the place of the infinite dimensional 
metric ${\bf g}$, can be determined by a number of physical
requirements. It has been shown in \cite{PJ} that this is possible at the 
leading order, when the Laplacian acts on local functionals. 
We will see here that this is the case also for the next order. 
Following similar steps with \cite{PJ} we
require that ${\bf G}$ is a regularisation of the inverse metric, so we 
will assume
that it depends on a cut-off parameter, $s$, with the dimensions of squared
length, and takes the form
$\label{eq:kernform}{\bf G}^{\mu _1\mu _2}_s(\sigma _1,\sigma _2)={\cal G}
_s(\sigma _1-\sigma _2)\,{\ K}^{\mu _1\mu _2}(\sigma _1,\sigma _2;s)$,
where ${\cal G}_s(\sigma _1-\sigma _2)\rightarrow \delta (\sigma _1-\sigma
_2)$ as $s\rightarrow 0$. ${\ K}$ is expandable as a power series in
positive integer powers of $s$, so that it has a finite limit as $s$ 
goes to
zero. Thus ${K}=\sum_{n=0}^\infty {K}_ns^n$ and ${K_0}^{\mu _1\mu 
_2}(\sigma
,\sigma )=g^{\mu _1\mu _2}(z(\sigma ))$ so that 
$\lim _{s\rightarrow 0}{\bf G}^{\mu _1\mu _2}(\sigma _1,\sigma _2)=
g^{\mu _1\mu _2}(\sigma _1)\delta (\sigma _1-\sigma _2)$.
To preserve the invariance of the theory under internal rotational symmetry
the kernel, ${\bf G}$, must be a second rank tensor under the 
restricted class of co-ordinate transformations of rigid rotations. 
We can define in addition to ${\bf g}^{\mu _1\mu _2}(\sigma _1,\sigma 
_2)$ its inverse as ${\bf g}
_{\mu _1\mu _2}(\sigma _1,\sigma _2)=g_{\mu _1\mu _2}(z(\sigma _1))\delta
(\sigma _1-\sigma _2)$ so their contraction gives
\begin{equation}
\int d\sigma {\bf g}_{\mu _1\mu }(\sigma _1,\sigma){\bf g}^{\mu \mu
_2}(\sigma ,\sigma _2)=\delta _{\mu _1}^{\mu _2}\delta (\sigma _1-\sigma
_2)\equiv {\bf I}_{\mu _1}^{\mu _2}(\sigma _1,\sigma _2) 
\end{equation}
where ${\bf I}_{\mu _1}^{\mu _2}(\sigma _1,\sigma _2)$ is the infinite
dimensional Kronecker delta and it is equal to the functional derivative 
of $ z^{\mu _2}(\sigma _2)$ with respect to $z^{\mu _1}(\sigma _1)$, so 
(\ref{co1}) can be rewritten more compactly as
$\left[ z^\mu (\sigma ),\pi _\nu (\sigma ^{\prime })\right] =i\alpha {\bf I}
_\nu ^\mu (\sigma ,\sigma ^{\prime })$.

The momentum operator we used in the commutation relation (\ref{momac}) 
has to be covariant. This property can be carried out to its functional 
differential operator representation. Given the
infinite dimensional metric we can follow the usual construction of the
Levi-Civita connection, ${\bf D}$, which will transform covariantly under
general co-ordinate transformations and therefore under our restricted
transformations. Thus if it acts on a scalar will reduce to the usual
functional derivative. For the spacial case of an infinite dimensional 
ultra-local vector ${\bf V}^{\mu_1}(\sigma _1)$ we get 
\begin{equation}
{\bf D}_{\mu _2}(\sigma _2){\bf V}^{\mu _1}(\sigma _1)={\frac{\delta {\bf V}
^{\mu _1}(\sigma _1)}{\delta z^{\mu _2}(\sigma _2)}}+\int d\sigma _3\,{\bf 
\Gamma }_{\mu _2\mu _3}^{\mu _1}(\sigma _1,\sigma _2,\sigma _3){\bf V}^{\mu
_3}(\sigma _3) =(D_{\mu _2}V^{\mu_1})|_{z(\sigma _1)}\,\delta (\sigma _1-\sigma _2)
\end{equation}
where the infinite dimensional Christoffel symbol is related to that on $S^N$
by 
${\bf \Gamma }_{\mu _2\mu _3}^{\mu _1}(\sigma _1,\sigma _2,$ $\sigma _3)=\delta
(\sigma _1-\sigma _2)\,\delta (\sigma _2-\sigma _3)\,{\ \Gamma }_{\mu _2\mu
_3}^{\mu _1}(z(\sigma _1))$
and $D$ is the covariant derivative on $S^N$. 
In addition, we can define a finite dimensional intrinsic derivative
${\cal D}={\partial /\partial \sigma}+z^{\prime \mu }D_\mu $, 
which is a total differential of the $\sigma$ variable with invariant 
transformation properties.

Since we work in a Hamiltonian formalism, Poincar\'e invariance is not
manifest and must be imposed by demanding that the generators of these
transformations satisfy the Poincar\'e algebra.
Ignoring regularisation the Poincar\'e generators are the Hamiltonian, 
given in 
(\ref {hami}) which generates time $\tau$ translations, the momentum $P=\int 
d\sigma z^{\prime \mu 
}{\bf D}_\lambda (\sigma )$ which generates space $\sigma$ translations and 
the Lorentz generator $L=-\alpha M+{\alpha}^{-1} N$, where 
\begin{equation}
\label{lorenzo}
M={\frac 12}\int d\sigma \,\sigma g^{\mu _1\mu _2}\,{\bf D}_{\mu _1}(\sigma )
{\bf D}_{\mu _2}(\sigma ),\quad N={\frac 12}\int d\sigma \,\sigma g_{\mu
_1\mu _2}z^{\prime \mu _1}z^{\prime \mu _2} 
\end{equation}
which generates Lorentz transformations in the $(\sigma,\tau)$ space.
Formally, operators $L$, $H$ and $P$ satisfy the Poincar\'e algebra 
$[P,H]=0$, $ [L,P]=H$, $[L,H]=P$. 
We require that this algebra holds for the regularised operators. The
momentum operator does not need to be regulated. We regulate the Laplacian
as in (\ref{lapal}) to yield a cut-off Hamiltonian $H_s$. As seen in 
\cite{PJ} the limit $s\rightarrow 0$ of the action of $H_s$ on $\Psi$ 
exists and represents the application of the Hamiltonian on $\Psi$.
Similarly the cut-off dependent Lorentz 
operator, $L_s$, should have a finite limit when applied to the physical
states. The commutator $[L,P]=H$ implies that $L$ should be regulated with
the same kernel as $H$, so we replace in (\ref{lorenzo}) the operator $M$ by 
\begin{equation}
\int d\sigma _1d\sigma _2\,{\frac{\sigma _1+\sigma _2}2}\,{\bf G}_s^{\mu
_1\mu _2}(\sigma _1,\sigma _2)\,{\bf D}_{\mu _1}(\sigma _1){\bf D}_{\mu
_2}(\sigma _2)\equiv M_s 
\end{equation}
The regularised versions of the Poincar\'e algebra impose 
conditions on the Kernel when acting on
local functionals named generally $F$. For example the regularised 
version of $([L,H]-P)F=0$ is 
\begin{equation}
\frac 14\left[ -\alpha M_{s}+{\alpha}^{-1}N,-\alpha\Delta 
_{s}+{\alpha}^{-1}V\right] F=PF 
\end{equation}
We demand that this equation holds order by order in $1/s$ up to order
zero. These are the terms that, in the absence of a regulator, involve two
functional derivatives at the same point on a single local functional. The
terms with positive powers of $s$ will disappear at the limit
$s\rightarrow 0$. They are equivalent to the $O(s^n)$ terms with $ n>0$
in the small-$s$ expansion (see \cite{Paul}), which is treated 
with the re-summation procedure in order to extract the desired zeroth
order term. Thus, by requiring $[M_s,\Delta_s]F=0 $, as a restriction for
the Kernel, and by ignoring the positive powers of $s$ in the intermediate
steps of the calculation we obtain a better approximation for the first
term of the small-$s$ expansion. Also we demand $M_sV=0$ 
and $\Delta_sN=0$.

We are interested in constructing the Schr\"odinger equation for
slowly varying fields. This allows us to expand 
the vacuum-functional in terms of local functionals. For the $O(N)$ 
$\sigma$-model they are integrals of functions of $\sigma$, $z(\sigma)$ 
and a finite number of its derivatives at the point $\sigma$. In order to 
construct the Kernel ${\bf G}$, we will consider the conditions that arise
from applying the regularised form of the Poincar\'e algebra to such test
functionals. It will be convenient to order them according to the powers 
of ${\cal D}$. If we consider the result of the action of the two functional
derivatives from $\Delta $ (or $M$) on a local test functional as a
differential operator of $\sigma $ acting on a delta function, then the
order of the operator, that is the highest number of covariant $\sigma $
derivatives acting on the delta function, depends on the highest number of
differentiations on the $z$'s used to construct the local functional. This
operator acting on one of the $\sigma $ arguments of the Kernel, via
integration by parts and setting its two arguments equal to each other, with
the application of the delta function, will demand the use of more terms 
of the Kernel expansion with respect to $s$, depending on the order of the 
operator and in conclusion on the order of the test functional.

In \cite{PJ} the least order local functionals were used, which have the form 
$F_n\equiv $ $\int d\sigma$ $f(z(\sigma ),\sigma )_{\mu _1...\mu _n}$ $
z^{\prime \mu _1}...z^{\prime \mu _n}$ where $f$ is ultra-local. 
Having in mind the transformation properties (rotation invariance) of ${\bf G
}$ and its dimension (inverse length) we can set 
$\,\,{\bf G}^{\mu \nu }(\sigma ,\sigma )=\frac 1{\sqrt{s}}b_0^0g^{\mu
\nu }$
so that 
\begin{equation}
\label{exp2}
\Big( \left( \left. {\cal D}\right| _\sigma +\left. {\cal D}
\right| _{\sigma ^{\prime }}\right){\bf G}^{\mu \nu }(\sigma ,\sigma ^{\prime
})\Big)  _{\sigma =\sigma ^{\prime }}={\cal D}\left( \frac {b_0^0}
{\sqrt{s}}g^{\mu \nu }\right) =0 
\end{equation}
up to zeroth order in $s$ and 
\begin{equation}
\label{exp3} 
\Big( \left. {\cal D}\right| _\sigma \left. {\cal D}
\right| _{\sigma ^{\prime }}{\bf G}^{\mu \nu }(\sigma ,\sigma ^{\prime
})\Big) _{\sigma =\sigma ^{\prime }}=-\frac 1{\sqrt{s}
^3}(b_0^1g^{\mu \nu }+sb_1^1g_{\lambda \rho }z^{\prime \lambda }z^{\prime
\rho }g^{\mu \nu }+sb_2^1z^{\prime \mu }z^{\prime \nu }) 
\end{equation}
where $b_1^0$, $b_0^1$, $b_1^1$, $b_2^1$, ... are dimensionless constants. $
b_0^0$ and $b_0^1$, given by $b_0^0=\sqrt{s}{\cal G}_s(0)$ and 
$b_0^1=\sqrt{s}^3{\cal G}^{\prime\prime }_s(0)$,
are determined by our choice of regularisation of the
delta-function ${\cal G}_s$.
As it was shown in \cite{PJ}, these coefficients can be determined by 
demanding the
closure of the Poincar\'e algebra. 
This is achieved by taking 
$b_1^1=-b_2^1=-b_0^0/a^2$,
with which we can built ${\bf G}$ up to first order.

In the same way we assume that for the next order the constraint
equations resulting from the closure of the Poincar\'e algebra, will be
independent of the general form of the test functional, as soon as the
differential operator is of order four. We can apply the Laplacian on $\int
d\sigma \,f(z(\sigma ),\sigma )_{\mu \nu }{\cal D}z^{\prime \mu }{\cal D}
z^{\prime \nu }\equiv F$, where $f$ is ultra-local. Then, the two functional
derivatives in the Laplacian will generate a fourth order differential
operator acting on $\delta (\sigma _1-\sigma _2)$. The consequence of this
is that ${\Delta _s}F$ will depend on the fourth derivative of the Kernel
evaluated at co-incident points. Demanding the closure of the Poincar\'e
algebra acting on $F$ will constrain this quantity. $F$ is the lowest order
functional of the general form $\int d\sigma f_{\rho _1...\rho _n}{\cal D}
z^{\prime \rho _1}...{\cal D}z^{\prime \rho _n}$ that gives a constrain to
this order.
To simplify the calculations we notice that the part of the result
of the action of the Lorentz operator on a general functional, which will
contribute to the commutation relations of the Poincar\'e algebra, is the
non-linear one in $\sigma$. 

We treat $F$ as a scalar so that 
\begin{equation}
{\bf D}_\mu (\sigma )F={\frac{\delta F}{\delta z^\mu (\sigma )}}=D_\mu
\,f_{\rho_1 \rho_2 }{\cal D}z^{\prime \rho_1 }{\cal D}z^{\prime \rho_2 }+2
{\cal D}^2(f_{\mu \rho_2 }{\cal D}z^{\prime \rho_2 })-2f_{\rho_1 \rho_2
}z^{\prime \kappa}z^{\prime \lambda}
R_{\mu \kappa \lambda}^{{\,\,\,\,\,\,\,\,}\rho_1 }
{\cal D}z^{\prime \rho_2 } 
\end{equation}
which is an infinite component co-vector. Using the commutators of ${\bf D}$ 
and ${\cal D}$, given in \cite{PJ}, we can show that
\begin{eqnarray}
\label{durh}
M_sF&=&\int d\sigma \sigma {\bf G}^{\mu \nu }(\sigma ,\sigma )\Big((D_\mu D_\nu
f_{\rho _1\rho _2}-2R_{\mu \rho _1 \nu }^{\,\,\,\,\,\,\,\,\,\,\,\,\lambda }\,f_{\lambda
\rho _2}){\cal D}z^{\prime \rho _1}{\cal D}z^{\prime \rho _2} +
\nonumber
\\
\nonumber
\\
&&
2f_{\rho _1\rho _2}R_{\nu \kappa_1\kappa_2}^{\,\,\,\,\,\,\,\,\,\,\,\,\,\,\rho _1}
R_{\mu \lambda_1 \lambda_2}^{\,\,\,\,\,\,\,\,\,\,\,\,\,\,\,\rho _2}
z^{\prime \kappa_1}z^{\prime \kappa_2}z^{\prime \lambda_1}z^{\prime \lambda_2}-
4D_\mu f_{\rho _1\rho _2}R_{\nu \kappa \lambda}^{\,\,\,\,\,\,\,\,\,\,\,\rho_1}
z^{\prime \kappa}z^{\prime \lambda}{\cal D}z^{\prime \rho _2}\Big)+ 
\nonumber
\\ 
\nonumber
\\
&&
4\int d\sigma \,\sigma \left( {\cal D}_\sigma ^2\,{\bf G}^{\mu \nu }(\sigma
,\sigma ^{\prime })\right) _{\sigma =\sigma ^{\prime }}(D_\nu f_{\mu \rho _2}
{\cal D}z^{\prime \rho _2}-f_{\rho_1 \mu }R_{\nu \kappa \lambda}
^{\,\,\,\,\,\,\,\,\,\rho_1} z^{\prime \kappa}z^{\prime \lambda})+ 
\nonumber
\\ 
\nonumber
\\
&&
2\int d\sigma \sigma \left( {\cal D}_\sigma ^4{\bf G}^{\mu \nu }(\sigma
,\sigma ^{\prime })\right) _{\sigma =\sigma ^{\prime }}f_{\mu \nu }+ 
4\int d\sigma \,{\bf G}^{\mu \nu }(\sigma ,\sigma )f_{\rho _1\rho _2}R_{\,\,\,\,\nu
\kappa \mu }^{\rho _1}z^{\prime \kappa}{\cal D}z^{\prime \rho _2} 
\end{eqnarray}
Using relations (\ref{exp2}), (\ref{exp3}) and the additional
\begin{eqnarray}
&&\left( {\cal D}_\sigma ^4{\bf G}^{\mu \nu }(\sigma ,\sigma ^{\prime
})\right) _{\sigma =\sigma ^{\prime }}={\frac 1{{\sqrt{s}}^5}}
\Big(\,b_0^2\,g^{\mu \nu }+sb_1^2\,g_{\lambda \rho }z^{\prime \lambda }z^{\prime
\rho }\,g^{\mu \nu }+sb_2^2\,z^{\prime \mu }z^{\prime \nu } 
\nonumber
\\
\nonumber
\\
&&
+s^2b_3^2\,(g_{\lambda \rho }z^{\prime \lambda }z^{\prime \rho })^2\,g^{\mu
\nu }+s^2b_4^2\,g_{\lambda \rho }z^{\prime \lambda }z^{\prime \rho
}\,z^{\prime \mu }z^{\prime \nu } 
+s^2b_5^2\,{\cal D}z^{\prime \mu }{\cal D}z^{\prime \nu
}+s^2b_6^2\,g_{\lambda \rho }{\cal D}z^{\prime \lambda }{\cal D}z^{\prime
\rho }\,g^{\mu \nu } 
\nonumber
\\
\nonumber
\\
&&
+s^2b_7^2\,z^{\prime (\mu }{\cal D}^2z^{\prime \nu )}+s^2b_8^2\,g_{\lambda
\rho }z^{\prime (\lambda }{\cal D}^2z^{\prime \rho )}g^{\mu \nu }\,\,\,\Big) , 
\end{eqnarray}
where $b_0^2=\sqrt{s}^5 {\cal G}^{\prime \prime \prime \prime}_s (0)$,
we can write $M_sF$ in terms of the constants $b_j^i$. Because of the
property 
\begin{equation}
\left( {\cal D}^k|_\sigma {\bf G}^{\mu \nu }(\sigma ,\sigma ^{\prime
})\right) _{\sigma =\sigma ^{\prime }}=-\left( {\cal D}^k|_{\sigma^{\prime }}
{\bf G}^{\mu \nu }(\sigma ,\sigma ^{\prime })\right) _{\sigma =\sigma
^{\prime }}+O(\sqrt{s}) 
\end{equation}
for $k$ an odd integer, and the symmetry of $\Delta _s$ and $M_s$ in $\sigma
,\sigma ^{\prime }$ there will not be any odd number of intrinsic
derivatives acting on ${\bf G}$ in our final expressions so we will not need
their expansion in terms of $b$'s.

After substituting into (\ref{durh}) we have
\begin{eqnarray}
&&
M_s\int d \sigma f_{\mu \nu }{\cal D}z^{\prime \mu }{\cal D}z^{\prime
\nu }= \int d\sigma \,\sigma \Big\{ {\frac 2{{\sqrt{s}}^5}}b_0^2f_\nu ^\nu + 
\nonumber
\\
\nonumber
\\
&&
{\frac 1{{\sqrt{s}}^3}}(4b_0^1D^\nu f_{\nu \mu}{\cal D}z^{\prime \mu}
-4b_0^1f_{\mu \nu }R_{\,\,\,\kappa \lambda}^{\mu \,\,\,\,\,\,\,\nu }
z^{\prime \kappa}z^{\prime \lambda}
+2b_1^2f_\nu ^\nu g_{\kappa \lambda}z^{\prime \kappa}z^{\prime \lambda}+
2b_2^2f_{\mu \nu} z^{\prime \mu }z^{\prime \nu })+ 
\nonumber
\\
\nonumber
\\
&&
{\frac 1{{\sqrt{s}}}}\Big( (J_2f)_{\mu \nu }{\cal D}z^{\prime \mu }{\cal D}z^{\prime \nu
}+(J_4f)_{\mu \nu \kappa \lambda }z^{\prime \mu }z^{\prime \nu }z^{\prime
\kappa }z^{\prime \lambda }+(J_3f)_{\mu \nu }z^{\prime (\mu }{\cal D}
^2z^{\prime \nu )}\Big) \Big\}+ 
\nonumber
\\
\nonumber
\\
&&
\int d\sigma \,{\frac 4{{\sqrt{s}}}}b_0^0{\frac{1-N}{a^2}}f_{\mu \nu
}z^{\prime \mu }{\cal D}z^{\prime \nu}
\end{eqnarray}
where
\begin{eqnarray}
(J_2f)_{\mu \nu }&=&b_0^0\widetilde{\Delta }f_{\mu \nu }-2b_0^0R_{(\mu
}^\rho f_{\nu )\rho }+2b_5^2f_{\mu \nu }+2b_6^2f_\kappa ^\kappa g_{\mu \nu } 
\\
\nonumber
\\
(J_3f)_{\mu \nu }&=&2b_7^2f_{\mu \nu }+2b_8^2f_\kappa ^\kappa g_{\mu
\nu } 
\\
\nonumber
\\
(J_4f)_{\mu \nu \kappa \lambda }&=&2(b_3^2-{\frac{b_0^0}{a^4}}
)f_\rho ^\rho g_{(\mu \nu }g_{\kappa \lambda )}+2(b_4^2+{\frac{b_0^0}{a
^4}})f_{(\mu \nu }g_{\kappa \lambda )} 
\end{eqnarray}
We can easily read of the action of $\Delta _s$ from above (term
linear in $\sigma $). In order to compute the commutator of $M_s$ and 
$\Delta _s$ on $\int f_{\mu \nu }{\cal D}z^{\prime \mu }{\cal D}z^{\prime \nu
}$, we need the following relations:
\begin{eqnarray}
\label{its}
\Delta _s\int f_{\kappa \lambda }z^{\prime \kappa }{\cal D}z^{\prime \lambda
}&=&\int d\sigma \Big({\frac 1{{\sqrt{s}}^3}}2b_0^1D^\lambda f_{\kappa \lambda }z^{\prime
\kappa }+
\nonumber
\\
\nonumber
\\
&&
{\frac 1{{\sqrt{s}}}}b_0^0(\tilde \Delta f_{\kappa \lambda }+2{\frac{1-N}{
a ^2}}f_{\kappa \lambda })z^{\prime \kappa }{\cal D}z^{\prime \lambda }\Big) 
\\
\nonumber
\\
M_s\int f_{\kappa \lambda }z^{\prime (\kappa }{\cal D}^2z^{\prime \lambda
)}&=&...+\int d\sigma \Big\{ {\frac 3{{\sqrt{s}}^3}}b_0^1D^\lambda f_{\kappa \lambda
}z^{\prime \kappa } +
\nonumber
\\
\nonumber
\\
&&\frac {b_0^0}{{\sqrt{s}}}\Big(D^\kappa f_{\kappa \lambda} 
{\cal D}^2z^{\prime \lambda}-
R_{\,\,\,\,\gamma \rho}^{\nu\,\,\,\,\,\,\,\,\lambda}
D_{\nu} f_{\kappa \lambda}z^{\prime \kappa}z^{\prime \rho}z^{\prime \gamma}+
\nonumber
\\
\nonumber
\\
&&\frac 3{a^2} f_\kappa ^\kappa g_{\gamma \rho}z^{\prime \gamma}
{\cal D}z^{\prime \rho}- \frac {3N}{a^2} f_{\kappa \lambda}z^{\prime \kappa}
{\cal D}z^{\prime \lambda} \Big) \Big\}
\\
\nonumber
\\
M_s\int f_\kappa {\cal D}z^{\prime \kappa }&=&...-\int d\sigma {\frac 
2{{\sqrt{s}}}}b_0^0R_\lambda ^\kappa f_\kappa z^{\prime \lambda }
\\
\nonumber
\\
M_s\int f_{\kappa \lambda }z^{\prime \kappa }z^{\prime \lambda }&=&...+
\int d\sigma {\frac 2{\sqrt{s}}}b_0^0D^\kappa f_{\kappa \lambda }z^{\prime \lambda } 
\label{nih}
\end{eqnarray}
where $...$ represents the linear in $\sigma$ part of the action of $M_s$ on the
specific functional.
Using relations (\ref{its})-(\ref{nih}) we derive the action of the 
operators $M_s\Delta_s$ 
and $\Delta _sM_s$ on $\int f_{\mu \nu }{\cal D}z^{\prime \mu }{\cal D}
z^{\prime \nu }$ to be
\begin{eqnarray}
\label{suse}
&&
M_s\Delta _s\int f_{\mu \nu }{\cal D}z^{\prime \mu }{\cal D}z^{\prime \nu }= 
\nonumber
\\
\nonumber
\\
&&
...+\int d\sigma \Big \{  {\frac 1s}\Big[4b_0^0{\frac{1-N}{a ^2}}\Big( b_0^0\tilde \Delta
f_{\mu \nu }-2b_0^0R_\mu ^\rho f_{\nu \rho }+2b_5^2f_{\mu \nu
}+2b_6^2f_\rho ^\rho g_{\mu \nu }\Big)z^{\prime \mu }{\cal D}z^{\prime \nu
}+ 
\nonumber
\\
\nonumber
\\
&&
4b_0^0D^\mu \Big(2(b_3^2-{\frac{b_0^0}{a ^4}})f_\rho ^\rho g_{(\mu \nu
}g_{\kappa \lambda )}+2(b_4^2+{\frac{b_0^0}{a ^4}})f_{(\mu \nu
}g_{\kappa \lambda )}\Big)z^{\prime \nu }z^{\prime \kappa }z^{\prime \lambda }
\nonumber
\\
\nonumber
\\
&&
b_0^0 D^\nu (J_3 f)_{\nu \lambda} 
{\cal D}^2z^{\prime \lambda}-b_0^0 \, R_{\,\,\,\gamma \rho}^
{\nu\,\,\,\,\,\,\,\lambda}
D_{\nu} (J_3 f)_{\kappa \lambda}z^{\prime \kappa}z^{\prime \rho}
z^{\prime \gamma}+
\nonumber
\\
\nonumber
\\
&&
\frac {3b_0^0}{a^2} (J_3 f)_\nu ^\nu g_{\gamma \rho}z^{\prime \gamma}
{\cal D}z^{\prime \rho}- \frac {3Nb_0^0}{a^2} 
(J_3f)_{\gamma \rho}z^{\prime \gamma}
{\cal D}z^{\prime \rho} \Big]
\nonumber
\\
\nonumber
\\
&&
{\frac 1{s^2}}\Big(4b_0^1b_0^02{\frac{1-N}{a ^2}}D^\nu f_{\nu \rho
}+4b_0^0b_1^2D_\rho f_\kappa ^\kappa +4b_0^0b_2^2D^\mu f_{\mu \rho }+ 
\nonumber
\\
\nonumber
\\
&&
6b_0^1b_7^2D^\nu f_{\nu \rho }+6b_0^1b_8^2D_\rho f_\kappa ^\kappa +4{\frac{
b_0^1b_0^0}{a ^2}}(2D_\rho f_\kappa ^\kappa -2D^\nu f_{\nu \rho
})\Big)z^{\prime \rho } \Big\} 
\end{eqnarray}
and
\begin{eqnarray}
\label{liz}
\Delta _sM_s\int f_{\mu \nu }{\cal D}z^{\prime \mu }{\cal D}z^{\prime \nu }&=& 
...+\int d\sigma \Big\{ {\frac 4s}{b_0^0}^2{\frac{1-N}{a ^2}}(\tilde \Delta
f_{\mu \nu }+2{\frac{1-N}{a ^2}}f_{\mu \nu })z^{\prime \mu }{\cal D}
z^{\prime \nu }
\nonumber
\\
\nonumber
\\
&&
+{\frac 4{s^2}}b_0^0{\frac{1-N}{a ^2}}2b_0^1D^\nu f_{\mu \nu
}z^{\prime \mu } \Big\}
\end{eqnarray}
The $O(1/s^3)$ order does not contribute to the 
commutator. 
The $O(1/s^2)$ order in (\ref{suse}) and (\ref{liz}) can be factorised with
coefficients the various $z$ combinations
which result the following relations
\begin{equation}
\label{astana}
-4b_0^1b_0^0{\frac 1{a ^2}}+2b_0^0b_2^2+3b_0^1b_7^2=0,\,\,\,\,\,\,\,
4\,b_0^1b_0^0{\frac 1{a ^2}}+2b_0^0b_1^2+3b_0^1b_8^2=0 
\end{equation}
The $O(1/s)$ order gives
$b_3^2=-b_4^2={{b_0^0}/{a ^4}}$ and $b_5^2=b_6^2=b_7^2=b_8^2=0$, 
so that (\ref{astana}) becomes
$-2b_0^1/{a ^2}+b_2^2=0$ and 
$2\,b_0^1/{a ^2}+b_1^2=0$.
From these relations the $b$'s needed up to this order in
the Kernel are completely determined. Assuming that the theory is
consistent these coefficients are the same for every test functional we use with the
same highest number of $\sigma $ derivatives, acting on $z$, within $F$. 
We can write $K$ as a Taylor expansion of the $\sigma$ variables with the help of 
${\bf W}^{\mu_1}_{\mu_2}(\sigma_1,\sigma_2)$, defined by 
${\cal D}|_{\sigma_1}{\bf W}^{\mu_1}
_{\mu_2}(\sigma_1,\sigma_2)=0,$ and ${\bf W}^{\mu_1}
_{\mu_2}(\sigma,\sigma)=\delta^{\mu_1}_{\mu_2}$ (see \cite{PJ}). Up to order 
$O\left((\sigma_1-\sigma_2)^5\right)$ we get
\bq
&&
K^{\mu_1\mu_2}(\sigma_1,\sigma_2)={\bf W}^{\mu_1}_\nu
(\sigma_1,\sigma_2)
\Big( g^{\nu\mu_2}|_{z(\sigma_2)}-{\textstyle
{1\over 2}}{(\sigma_1-\sigma_2)^2}
R^{\nu\,\,\,\mu_2}_{\,\,\,\lambda\,\,\,\rho}
z^{\prime\lambda}
z^{\prime\rho}|_{z(\sigma_2)}+
\nonumber\\
&&
{1\over {4!(N-1)}}{(\sigma_1-\sigma_2)^4}{R_{\kappa \sigma}
z^{\prime\kappa}z^{\prime\sigma}
R^{\nu\,\,\,\mu_2}_{\,\,\,\lambda\,\,\,\rho}z^{\prime\lambda}z^{\prime\rho}
|_{z(\sigma_2)}}+
\nonumber\\
&&
{b_0^1 \over {6\,b_0^0}} {1\over s}{(\sigma_1-\sigma_2)^4}
{R^{\nu\,\,\,\mu_2}_{\,\,\,\lambda\,\,\,\rho}z^{\prime\lambda}z^{\prime\rho}
|_{z(\sigma_2)}}
\Big)
\eq
While up to order $O\left((\sigma_1-\sigma_2)^3\right)$, $K$ is $s$ independent, an 
$s$ dependent term appears in the next order. In a similar way we can 
determine more terms of the expansion by asking the closure of the Poincar\'e 
algebra when acting on test functionals of higher order.


\end{document}